# A Low-Temperature Specific Heat Study of the Giant Dielectric Constant Materials

C. P. Sun[1], Jianjun Liu[2, 5], J. -Y. Lin[3], Chun-gang Duan[4, 5], W. N. Mei[2, 5], and H. D. Yang[1, *]

[1]*Department of Physics, National Sun Yat-Sen University, Kaoshiung 804, Taiwan, Republic of China*
[2]*Department of Physics, University of Nebraska at Omaha, Nebraska, 68182-0266, USA*
[3]*Institute of Physics, National Chiao-Tung University, Hsinchu 300, Taiwan, Republic of China*
[4]*Department of Physics, University of Nebraska-Lincoln, Lincoln, Nebraska, 68588, USA*
[5]*Center for Material Research and Analysis, University of Nebraska-Lincoln, Lincoln, Nebraska, 68588, USA*

Low-temperature specific-heat study has been performed on the insulating giant dielectric constant material $CaCu_3Ti_4O_{12}$ and two related compounds, $Bi_{2/3}Cu_3Ti_4O_{12}$ and $La_{0.5}Na_{0.5}Cu_3Ti_4O_{12}$, from 0.6 to 10 K. From analyzing the specific heat data at very low-temperature range, 0.6 to 1.5 K, and moderately low-temperature range, 1.5 to 5 K, in addition to the expected Debye terms, we noticed significant contributions originated from the linear and Einstein terms, which we attributed as the manifestation of low-lying elementary excitations due to lattice vibrations occurred at the grain boundaries and induced by local defects. Together with the findings on electronic and mechanical properties, a phenomenological model is proposed to explain the high dielectric constant behaviors at both low and high frequency regions.

PACS numbers: 65.40.Ba, 65.40 –b, 77.22-d.

Recently $CaCu_3Ti_4O_{12}$ (CCTO) and related compounds have attracted much attention because of its extra ordinary dielectric behavior. The ceramic samples were first found to have a gigantic dielectric constant (K, up to $10^4$) independent of frequency and temperature in the range of DC - $10^5$ Hz and 100 - 600 K, then dropped steadily down to about 100 at $10^6$ Hz and maintained its constant value until rather high frequency ~$10^9$ Hz [1, 2]. Later, a giant value of $10^5$ was reported in a CCTO single crystal; yet twin and domains still existed [3]. Usually, a large dielectric constant results from atomic displacements in a non-centrosymmetrical structure near its phase transition temperature, e.g. in a ferroelectric. However, CCTO has a body-centered cubic perovskite related structure and there are no anomalies in crystal structure and lattice vibrations down to 35 K [4, 5]. In addition, the values of the dielectric constant varied sensitively on preparation methods and post treatments of the ceramic samples [6-7]. Also, the dielectric constant of a thin-film single crystal sample only had a value of about 100 [8]. Therefore, theoretical calculations [9-10] and many experimental results suggested that the large dielectric constant of CCTO unlikely originates from its intrinsic property but from microstructures of the ceramic sample or a single crystal. It has been suggested that the high dielectric constant could be due to creation of barrier layer capacitances at twin boundaries [1, 3] or at the interfaces between grains and grain-boundaries [11-12] or between the sample and the electrodes [2, 13]. Current-voltage measurements show that the grain boundaries in the ceramic sample play an important role in the CCTO electrical properties [14]. We thoroughly studied the temperature dependence of the permittivity and impedance of CCTO and related compounds such as $Bi_{2/3}Cu_3Ti_4O_{12}$, $La_{2/3}Cu_3Ti_4O_{12}$, and $Y_{2/3}Cu_3Ti_4O_{12}$ [12, 15]. We found that all the compounds have similar dielectric and electric behaviors, although they have different dielectric constants. In all the compounds we detected a Debye-like relaxation in the permittivity formalism with a dielectric constant independent of the temperature and frequency in a wide range. We then observed two electric responses with very different intensity and response frequencies in impedance formalism. Based on these results, we suggested a *shell-core model that the ceramic samples consist of semiconducting grains partitioned from each other by poorly conducting grain boundaries* (see SEM image in figure 1), and ascribed these two electrical responses in the impedance formalism to the grains and grain-boundaries, respectively. The detected Debye-like relaxation and large dielectric constant originated from the Maxwell-Wagner relaxation at the interfaces between grains and grain -boundaries. Besides, we used X-ray diffraction and high-pressure techniques [16], and found that ceramic CCTO samples behaved differently under hydrostatic and uniaxial compression conditions. Recently, a study [17] utilizing both x-ray photoelectron (XPS) spectroscopy and energy dispersive x-ray spectroscopy (EDX) provided a strong support to this possible grain-boundary structure, and a new giant dielectric constant material $K_xTi_yNi_{1-x-y}O$ is also well described by the insulating grain boundary model. [18]

In this letter, low-temperature specific-heat (LTSH) measurement of CCTO and two other high-K compounds, $Bi_{2/3}Cu_3Ti_4O_{12}$ and $La_{0.5}Na_{0.5}Cu_3Ti_4O_{12}$, one with defects and the other is a mixture, were performed from 0.6 to 10 K. This is a well-developed technique known to provide information about the nature of the ground state according to different characteristics of the elementary excitations. Specific-heat results can be classified into independent contributions, such as free electron, lattice vibration, magnetic ordering, etc, in which the electronic part is proportional to $T$. More

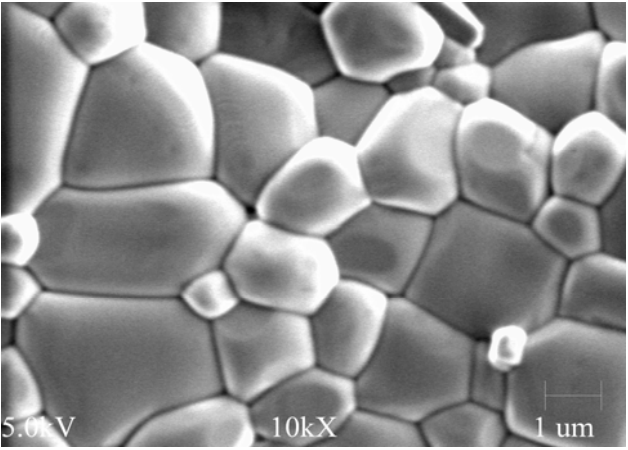

Figure 1. SEM image of CCTO sample. Notice the grains are of the size of few μm, determined by using a JOELJSM840A scanning electron microscope set at 5 kV. The ample surfaces were gold-coated prior examination.

-over, lattice term can be derived from the dispersion relations of lattice vibrations, which is usually separated into two parts: Debye and Einstein mode [17, 19-21].

Polycrystalline samples of CCTO, $Bi_{2/3}Cu_3Ti_4O_{12}$ and $La_{0.5}Na_{0.5}Cu_3Ti_4O_{12}$, were prepared by heating mixed stoichiometric amounts of the chemicals $CaCO_3$, $CuO$, and $TiO_2$ (or $Na_2CO_3$, $La_2O_3$, $Bi_2O_3$ when fabricating other compounds) at 1000 C for 30 hours with intermediate grinding. The final samples were ground into powder and verified to be in a single phase by Scintag PADV powder X-ray diffractometer. The LTSH $C(T)$ was measured by using a $^3$He heat-pulsed thermal relaxation calorimeter [19 - 21] whose essential method to calculate the heat capacity is to reach the thermal equilibrium of the sample and background less than a characteristic time (time constant $\tau$) subsequent to heat pulse is applied. After subtracting the addenda contribution, determined from a separate measurement, the remaining value is the heat capacity of sample presented in figure 2 by $C/T^3$ vs. $T$. In order to investigate the ground state properties of these high-K compounds, we analyze the LTSH data in the studied temperature range by using the fitting equation $C(T)=C_{hyp}+C_{linear}+C_{lattice}$ [22], where $C_{hyp}$ is the hyperfine contribution ($\sim 1/T^2$) caused by the tail of a Schottky anomaly produced by the interaction of the nuclear quadrupole moment with electric-filed gradient in the crystal. This usually takes place at the lowest temperature region. $C_{linear}$ is proportional to $T$, normally originated from the conduction electrons, and $C_{lattice}=C_D+C_E$ corresponds to Debye and Einstein modes, respectively. $C_D = \beta T^3 + \delta T^5$ includes linear frequency dispersion at low energy and anharmonic term of lattice vibration. β is related to a fundamental physical parameter, Debye temperature β $=1.944*10^6*n/\Theta_D^3$ where n is the number of atom per formula unit. It should be emphasized that the Debye lattice vibration term might contain the contribution of the antiferromagnetic magnon with $T_N$ ~25K, because both of them have the same temperature dependence of specific heat. Moreover, $C_E$ originates from low-lying optical modes, which are dispersionless, described as $C_D=D(T_E/T)^2\exp(T_E/T)/[\exp(T_E/T)-1]^2$, where $D = 3n_E R$, $n_E$ is the number of free Einstein-like modes per formula unit and $R$ is the gas constant, and $T_E$ is the characteristic temperature of Einstein mode. All the acceptable fits are listed in table 1, all the parameters are positive, within reasonable range and the standard deviations (S. D.) are small.

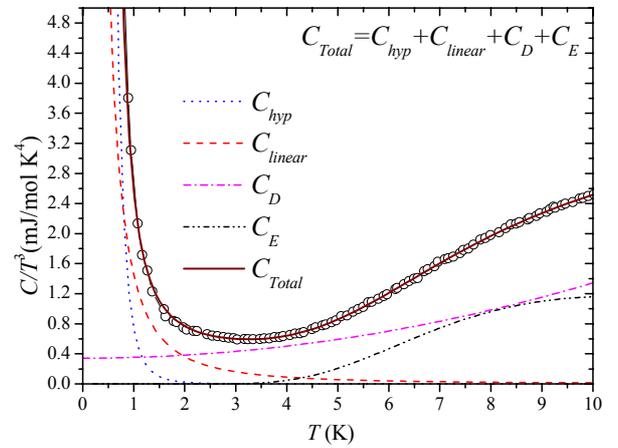

Figure 2. Low-temperature specific-heat data (black circles), plotted as $C(T)/T^3$ vs. $T$, from 0.6 to 10 K, and the best fit contains all the contributions is the solid line (S. D.=0.29).

We present our fitting results in table 1 and 2. Since three samples share the similar characteristics, we focus on the CCTO results. Here we found two interesting features: (1) the occurrence of Einstein modes at a low temperature range $T<$ 5K and (2) the manifestation of linear term in this good insulator (~0.45 MΩcm at RT [11], similar for other samples). Starting with the analysis, it was impossible to omit $C_E$ and $C_{linear}$. In fact the standard deviation for the best fit to be 16.30, which we could obtain without incorporating linear and Einstein terms. But after we included $C_E$, S. D. decreased immediately to 0.46. Then by adding $C_{linear}$, values of the standard deviation dropped to the neighbourhood of 0.3. In figure 2, the entire low-temperature (0.6 to 10 K) specific-heat data is shown together with the best fits, including all the contributions. Later, in order to evaluate the relative merits between $C_{hyp}$, $C_D$ and $C_E$ contributions, we present the data as following: in figure 3, we notice that in the plot of $C(T)T^2$ versus $T^3$ at the lowest temperature region, that is from 0.6 to 1.5 K, we observe clearly that $C_{hyp}+$

$C_{linear}$ describes the data better than $C_{hyp} + C_D$ (cubic term only). Finally, when combining the three terms, we reached a satisfactory agreement. Then, in figure 4, we plotted $C(T)/T$ versus $T^2$ from 1.5 to 5 K. It should be emphasized $C_E$ started to rise above the $C_{linear}+C_D$ around 3 to 4 K. Thus, from the above analysis, we deduce that there are other excitations on top of the Debye contribution raised from the conventional acoustical modes. The first stage is that the linear term known to affiliate with the conduction electrons; nevertheless, CCTO is identified to be a good insulator. Similar phenomena were reported in $C_{60}$ measurements [23], which were attributed to the disordered induced localized state but not a special feature in its excitation spectrum. In the present case, we associate this linear dependent feature in low-temperature specific heat with the *corrugated phonon modes* located on the grain boundaries with an elementary excitation spectrum, different from the conventional acoustical phonon modes behaving as $\omega \sim k^2$, where $\omega$ and $k$ are the frequency and wave vector, respectively. This particular excitation was used to portray bending motion of long-chain molecules [24] and bond angle vibration in the Cu-O plane of high-$T_c$ superconductors [25]; we found it commensurate with our earlier studies on various properties of CCTO [12, 15, 16]. We deduced that the samples are composed of grains with tough shells which can be regarded as quasi-two dimensional lattices, hence the corrugated phonon modes, with dispersion relation $\omega \sim k$, constitute the linear $T$ term in the specific-heat data. In addition, the scenario that the existence of a stiff insulating grain boundary, which is capable of storing large amounts of charges, facilitates the giant dielectric constant at low frequency region agreed with several existing theories [11, 12, 14].

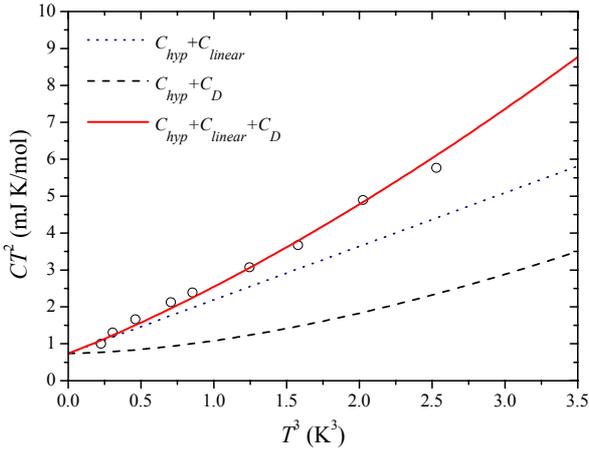

Figure 3. Specific-heat data, plotted as $C(T)T^2$ vs. $T^3$, from 0.6 to 1.5 °K, and different fits: the model contains (i) $C_{hyp}$ and $C_{linear}$ is the dotted line, (ii) $C_{hyp}$ and $C_D$ is the dashed line, and (iii) $C_{hyp}$, $C_{linear}$, and $C_D$ is the solid line.

Usually, the Einstein term is related to the low-lying optical modes, according to the previous detail studies [9, 10]. Optical mode frequencies for the cubic phase CCTO ranged from 125 to 560 (1/cm), so they are much higher than those observed in our specific-heat data, in which the Einstein term started to appear about 3 to 4 K, around 2 to 5 (1/cm). Thus, these modes are not generated from lattice vibrations of a perfect solid. Rather, they are expected to come up from local defects at the grain boundaries and/or caused by oxygen deficiency inside the grains, which are produced during the heating process. On the other hand, the existence of these low-lying optical modes may assist us to unravel the mechanism of that reasonably high dielectric constant, about 100 and higher, at higher frequency region ($10^6$ to $10^9$ Hz). In fact, sub-IR frequency dielectric constant was estimated to be around 50 to 70 by utilizing the effective charges, static dielectric constant, and frequencies of optical modes calculated from lattice dynamics [9, 10]. Thus, it is conceivable that the value will be enhanced when we include these low-lying optical modes. To further investigate the proposed mechanism, specific heat of high-K materials $La_{0.5}Na_{0.5}Cu_3Ti_4O_{12}$ and $Bi_{2/3}Cu_3Ti_4O_{12}$ were studied and compared with that of CCTO in the table 2. We notice that contributions from the linear term are prominent in all the samples; CCTO is the smallest among them. This may imply that its grain boundary shell is the thinnest, and their Einstein temperatures $T_E$ are of the same order of magnitude. We encourage more theoretical work to address these unsettled questions and determine if our observations are relevant to the giant dielectric constant phenomena in CCTO and related compounds.

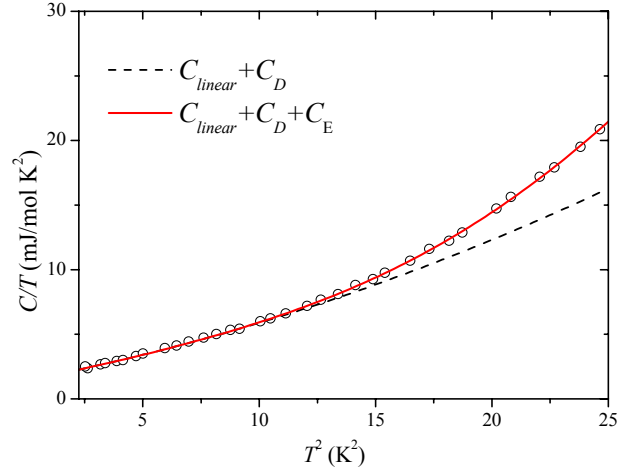

Figure 4. Specific-heat data, plotted as $C(T)/T$ vs. $T^2$, from 1.5 to 5 K, and different fits: the model contains (i) $C_{linear}$ and $C_D$ terms are the dashed line, (ii) $C_{linear}$, $C_D$ and $C_E$ terms are the solid line.

In conclusion, we present low-temperature specific heat on the giant dielectric constant material CCTO from 0.6 to 10 K. Clear indications of linear and Einstein terms contributions were observed on top of the commonly detected Debye term that originated from the conventional acoustic phonons. We examined the data at different ranges, 0.6-1.5 K and 1.5 to 5 K separately to emphasize their contribution. Similar results were obtained when we performed the analysis to $La_{0.5}Na_{0.5}Cu_3Ti_4O_{12}$, $Bi_{2/3}Cu_3Ti_4O_{12}$. Two distinct batches of CCTO polycrystalline sample, which have different oxygen pressure and baking temperatures, turned out only few percent changes in the fitting coefficients. Combined with the available theoretical and experimental studies on the electronic and mechanical properties of CCTO and related compounds, we present a phenomenological model that is consistent with many current theories.

Table 1: Fitting parameters for the low temperature specific heat data for $CaCu_3Ti_4O_{12}$. Equation we used is $C(T)=C_{hyp}+C_{linear}+C_{lattice}=A/T^2+\Lambda T+ \beta T^3+ \delta T^5+D(T_E/T)^2\exp(T_E/T)/[\exp(T_E/T)-1]^2$. (S.D.) and the values in the parentheses are the standard deviation and the error bar of the fit, respectively.

| A (mJ K/mol) | $\Lambda$ (mJ/mol K$^2$) | $\beta$ (mJ/mol K$^4$) | $\delta$ (mJ/mol K$^6$) | D (J/mol K) | $T_E$ (K) | S.D. |
|---|---|---|---|---|---|---|
| 1.2 (0.64) | 0 | 0.559 (0.031) | 0.0206(0.0004) | 0 | 0 | 16.30 |
| 1.27(0.10) | 0 | 0.489(0.016) | 0.0082(0.0005) | 9.0(0.5) | 53.8(0.4) | 0.46 |
| 0 | 2.32(0.19) | 0.259(0.025) | 0.0110(0.0005) | 7.0(0.4) | 50.7(0.5) | 0.43 |

Table 2: Best fitting parameters (same formula as table 1) for the three high-K compounds.

| $ACu_3Ti_4O_{12}$ | $\epsilon$ (f=1Hz) | A (mJ K/mol) | $\Lambda$ (mJ/mol K$^2$) | $\beta$ (mJ/mol K$^4$) | $\delta$ (mJ/mol K$^6$) | D (J/mol K) | $T_E$ (K) | S.D. |
|---|---|---|---|---|---|---|---|---|
| $CaCu_3Ti_4O_{12}$ | 9340 | 0.74(0.11) | 1.45(0.20) | 0.343(0.024) | 0.0100(0.0005) | 7.6(0.4) | 51.8(0.5) | 0.29 |
| $La_{0.5}Na_{0.5}Cu_3Ti_4O_{12}$ | 3560 | 1.63(0.23) | 6.67(0.37) | 0.906(0.054) | 0.0128(0.0007) | 6.7(0.3) | 43.2(0.4) | 1.35 |
| $Bi_{2/3}Cu_3Ti_4O_{12}$ | 2150 | 0 | 22.85(0.36) | 2.730(0.060) | 0.0060(0.0005) | 4.2(0.1) | 39.5(0.4) | 1.39 |

This work was partially supported by the National Science Council of Republic of China under contract NSC95-2112-M110-023. The Nebraska Research Initiative supports the authors at University of Nebraska.
*Corresponding author e-mail: yang@mail.phys.nsysu.edu.tw